\documentclass[twocolumn,tighten]{aastex6}
\usepackage{graphicx}

\newcommand{\tb}{\ensuremath{T_{\mathrm{B}}}}                      
\newcommand{\mujyb}{\,\mbox{\ensuremath{\mathrm{\mu Jy/beam}}}}  
\newcommand{\mjyb}{\,\mbox{\ensuremath{\mathrm{mJy/beam}}}}       
\newcommand{\lunits}{\,\ensuremath{\mathrm{erg\,s^{-1}\,Hz^{-1}}}}
\newcommand{\mjy}{\,\mbox{\ensuremath{\mathrm{mJy}}}}          
\newcommand{\msun}{\,\mbox{\ensuremath{\mathrm{M_{\odot}}}}}
\newcommand{\fdeg}{\ensuremath{.\!\!{\mathrm{\degr}}}} 
\newcommand{\myemail}{\,cristina.romero.fdi@mail.udp.cl}
\begin{document}

\title{The TDE ASASSN-14li and its host resolved at parsec scales with the EVN}

\author{Cristina Romero-Ca\~nizales\altaffilmark{1,2}$^{\star}$,
Jos\'e L. Prieto\altaffilmark{2,1}, Xian Chen\altaffilmark{3}, Christopher S. Kochanek\altaffilmark{4,5}, 
Subo Dong\altaffilmark{6}, Thomas W.-S. Holoien\altaffilmark{4,5,8}, Krzysztof Z. Stanek\altaffilmark{4,5} \&
Fukun Liu\altaffilmark{7}} 
\altaffiltext{1}{Millennium Institute of Astrophysics, Chile}
\altaffiltext{2}{N\'ucleo de Astronom\'{\i}a de la Facultad de Ingenier\'{\i}a, 
Universidad Diego Portales, Av. Ej\'ercito 441, Santiago, Chile}
\altaffiltext{3}{Instituto de Astrof\'{\i}sica, Facultad de F\'{\i}sica, Pontificia 
Universidad Cat\'olica de Chile, 7820436 Santiago, Chile}
\altaffiltext{4}{Department of Astronomy, Ohio State University, 140 West
18th Avenue, Columbus, OH 43210, USA}
\altaffiltext{5}{Center for Cosmology and AstroParticle Physics, The Ohio
State University, 191 W. Woodruff Ave., Columbus, OH 43210, USA}
\altaffiltext{6}{Kavli Institute for Astronomy and Astrophysics, Peking University, Yi
He Yuan Road 5, Hai Dian District, Beijing 100871, China}
\altaffiltext{7}{Department of Astronomy, Peking University, Yiheyuan Road 5, Haidian District,
Beijing 100871, China}
\altaffiltext{8}{US Department of Energy Computational Science Graduate Fellow}
\email{$^{\star}$ \myemail}

\begin{abstract}

We report European Very Long Baseline Interferometry Network (EVN) radio continuum 
observations of ASASSN-14li, one of the best studied tidal disruption events (TDEs) 
to date. At 1.7\,GHz with $\simeq12\times6$\,mas resolution, the emission is 
unresolved. At 5.0\,GHz with $\simeq3\times2$\,mas resolution, the radio emission 
shows an extended structure that can be modeled with two components: a core-like 
component and a fainter, possibly elongated source 4.3\,mas ($\sim2$\,pc) away.
Our observations are not conclusive as to the nature of the components, but 
three scenarios are discussed. One possibility is a core-jet/outflow morphology, 
thus making of ASASSN-14li the first TDE jet/outflow directly imaged. For this case, 
the projected separation between the two components can only be explained by superluminal 
motion, rather than the lower velocities inferred from low-resolution radio observations. 
However, typical fast moving jets have brightness temperatures $\sim5$ orders of magnitude 
higher than we find, thus making this scenario less likely. The second possibility is
that we are imaging a non-relativistic jet from past AGN/TDE activity. In this case
a past TDE is preferred given that the spatial extension and radio luminosity of the 
elongated component are consistent with the theoretical predictions for a TDE outflow.
Alternatively, the two sources could indicate the presence of a binary black hole, which
would then naturally explain the enhanced TDE rates of post-starburst galaxies.  
Future EVN observations will help us to distinguish between these scenarios.

\end{abstract}


\section{Introduction} \label{sec:intro}

Supermassive black holes (SMBHs) have masses between $10^6$--$10^{9.5}$\,\msun{} and they 
can be as luminous as an entire galaxy \citep[$>10^{45}$\,erg\,s$^{-1}$;][]{kormendy13}.
Their luminosity is regulated by the accretion of surrounding material. It has long been 
recognized that accretion onto an SMBH could operate in two modes: steadily or 
intermittently \citep{shields78}. Steady accretion can be sustained by hot gas from the host 
halo, stellar winds in the galactic nucleus, etc. \citep[][]{shull83}. Sudden/intermitent 
accretion can occur when the SMBH swallows clouds of cold gas \citep{tremblay16} or 
orbiting stars in their vicinity \citep{hills75,rees88,evans89}. In the latter of these feeding 
possibilities, the so-called tidal disruption events (TDEs), the total or partial disruption of 
the star provides fuel that is promptly fed to the accretion disk. 

TDEs have been identified at optical, UV and X-ray wavelengths, but little is known 
about their radio emission \citep[see the review by][]{komossa15}. It seems likely 
that only a small fraction of TDEs produce significant radio emission \citep[see, 
e.g.,][]{generozov16}. Either most TDEs do not launch jets, or their opening angles 
are so small that they easily go undetected \citep{bower13,vanvelzen13,generozov16}. 
High-resolution observations of NGC\,4845, the nearest TDE host, will likely 
provide valuable constraints on the TDE production of radio jets \citep{irwin15}.
 
{\it Swift}\,J$164449.3+573451$ is a particularly interesting case as radio and X-ray 
observations provided strong evidence of relativistic outflows from a TDE for the first 
time  \citep{bloom11,burrows11,levan11,zauderer11}. However, {\it Swift}\,J$164449.3+573451$ 
was observed using Very Long Baseline Interferometry (VLBI) repeatedly during 2011 
\citep{berger12} and the radio source remained unresolved at a resolution of 0.2\,mas, 
implying an upper limit on the expansion velocity of $\la3.8$\,$c$. Continued monitoring 
with the European VLBI Network (EVN) from 2011 to 2015 has shown that the radio source is 
still compact, and the measurements are in agreement with either a rapidly decelerating 
source (average $\beta\la0.3$\,$c$) or a very small viewing angle \citep{yang16}. 

A very important discovery was made with ASASSN-14li. Initially identified in 
the optical, this is one of the few TDEs to show evidence of both thermal and non-thermal 
components. ASASSN-14li was discovered by the All-Sky Automated Survey for Supernovae 
\citep[ASAS-SN;][]{shappee14} on 2014 November 22 at the center of the nearby ($z=0.0206$, 
$D=90.3$\,Mpc) post-starburst galaxy PGC\,043234 \citep{holoien16}. PGC\,043234 appears to 
be the remnant of a recent merger that likely hosted a low-luminosity Type II AGN prior to 
the TDE \citep{prieto16}. \citet{holoien16} present the optical, UV and X-ray properties of 
the event, followed by additional X-ray \citep{miller15,brown16}, UV \citep{cenko16}, radio 
\citep[][and this work]{alexander16,vanvelzen16}, and mid-IR \citep{jiang16} studies.

The radio observations by \citet{alexander16} pointed to a non-relativistic outflow 
that would have been ejected between 2014 August 11 and 25 August with an apparent velocity 
of $0.04c$--$0.12c$. The high-frequency radio observations by \citet{vanvelzen16} also 
favored a non-relativistic jet interpretation. \citet{krolik16} argue that the radio 
emission was produced by the ejected, unbound tidal debris.

Here, we report on EVN observations of ASASSN-14li from 2015. We 
describe the observations and their analysis in \S \ref{sec:obs}. 
We compare our observations with pre- and other post-TDE radio 
observations and discuss the implications of our results in \S 
\ref{sec:nat}. Our conclusions are presented in \S \ref{sec:sum}.

\section{Observations and data analysis} \label{sec:obs}

ASASSN-14li was observed with the EVN in 2015 under program EP096 (PI: J.\,L. 
Prieto). The observations consisted of three observing segments (see Table 
\ref{tab:evnlog}). Segment A was used to confirm the detection of compact 
radio emission. Subsequent dual-frequency observations (segments B and C) were 
made quasi-simultaneously to obtain an accurate spectral slope. Segment A lasted 
a total of 7\,hr and segments B and C lasted 5\,hr. Our observing setup resulted 
in an aggregate bit rate of 1024\,Mbps at 2 bit sampling. We used an integration 
time of 2\,s and 8 sub-bands of 16\,MHz each with full polarization. OQ208 and 
J1159$+$2914 were used as fringe finders. J1250$+$1621 was used as a phase calibrator.
This is a compact source at a projected distance of 1\fdeg5 from ASASSN-14li, thus
well below the maximum calibrator-to-target separation of 5\degr{} recommended for 
VLBI observations \citep[e.g.,][]{fomalont05,martividal10}. J1159$+$2914 remained 
constant at a 1$\sigma$ level in both time and frequency, and had a flux density of 
0.19$\pm$0.01\,Jy in segment B. 

\begin{figure*}[htb!]
\includegraphics[scale=0.65]{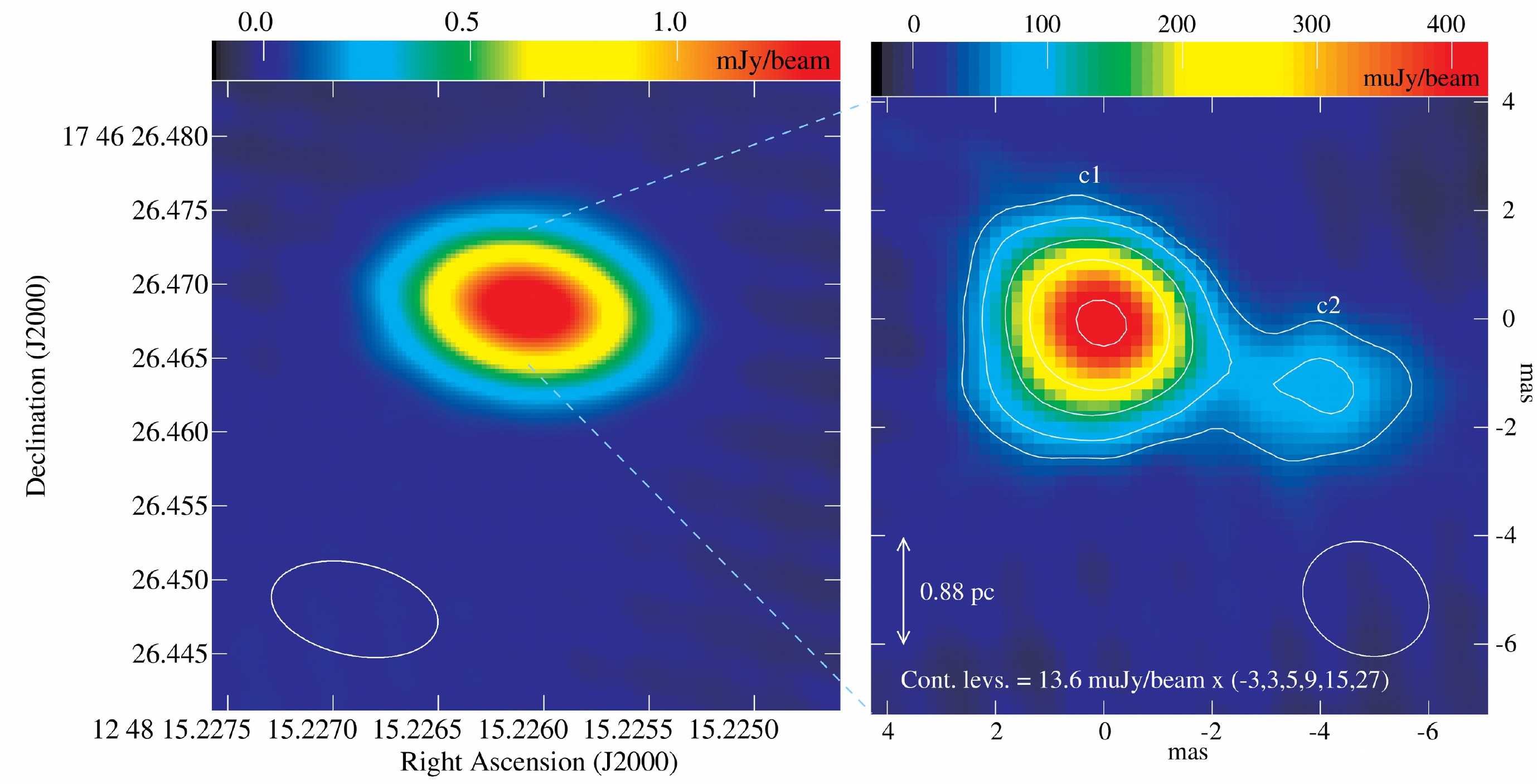}
\caption{EVN color maps of ASASSN-14li at 1.7\,GHz (left) and 5.0\,GHz (right). The ellipses 
at the bottom of each panel correspond to their convolving beams. The reference coordinates 
in the right panel are $\alpha(\mathrm{J}2000) = 12^{\mathrm{h}}48^{\mathrm{m}}15\fs2261$, 
$\delta(\mathrm{J}2000) = 17\degr46\arcmin26\farcs469$ (1\,mas$=0.44$\,pc). Labels correspond 
to the core-like (c1) and elongated (c2) components described in the text. \label{fig:ep096abc_img}}
\end{figure*}

\floattable
\begin{deluxetable}{cccclc}
\tablecaption{Log of the EP096 EVN observations. \label{tab:evnlog}}
\tablecolumns{6}
\tablewidth{0pt}
\tablehead{
\colhead{Segment} &  \colhead{Observing} &  \colhead{$\nu$} & \colhead{Time on}
 & \colhead{EVN stations\tablenotemark{a}} \\
\colhead{} & \colhead{date} & \colhead{(GHz)} & \colhead{source (hr)} & \colhead{}
}
\startdata
A & 2015 Apr 14 & 1.7 & 3.7 & Ef, Jb1, On, Hh, Sh, Wb, Tr, Jd, T6  \\
B & 2015 Jun 10 & 5.0 & 3.3 & Ef, Jb2, On, Hh, Sh, Wb, Tr, Mc, Sv, Zc, Bd, Nt, Ys \\
C & 2015 Jun 12 & 1.7 & 3.2 & Ef, Jb1, On, Hh, Sh, Wb, Tr, Mc, Sv, Zc, Bd \\
\enddata
\tablenotetext{a}{Locations and diameters are found at \url{http://www.evlbi.org/user_guide/EVNstatus.txt}}
\end{deluxetable}

We reduced the data within the National Radio Astronomy Observatory (NRAO) Astronomical 
Image Processing System ({\sc aips}). The calibration products from the EVN pipeline were 
used to inspect the quality of the data. We took into account ionospheric corrections and 
radio frequency interference to improve the calibration. Images of the calibrators 
revealed compact morphologies, so there was no need to add a model for fringe fitting. 

Figure \ref{fig:ep096abc_img} shows a color image of the naturally weighted maps from the 
2015 June observations at the two bands, and Table \ref{tab:ep096abc_param} 
reports the source model parameters for all the maps. The uncertainties in peak intensity 
and flux density are obtained by adding in quadrature the noise (rms) in each map 
to a 5\% uncertainty in the point-source calibration. For the flux density uncertainties, 
the noise term includes a factor equal to the number of beams covering the emitting region.
 
The images at 1.7\,GHz show an unresolved source regardless of the weighting scheme used 
for imaging in both segments A and C. For the observations in 2015 June, we made a 5.0\,GHz 
map using the same convolving beam as the one obtained at 1.7\,GHz with natural weighting, 
thus effectively degrading its resolution. This allowed us to obtain a global spectral index 
($\alpha$) of $-0.7\pm0.1$ (defined as $S_{\nu} \propto \nu^{\alpha}$). 

The natural weighted map at 5.0\,GHz (Figure \ref{fig:ep096abc_img}, right panel) 
shows an extended structure that can be modeled with two components: a core-like 
one, and a possibly elongated one with a $\sim5\sigma$ peak. Their ICRF2 J2000 peak 
positions are $\alpha_{\mathrm{c1}} = 12^{\mathrm{h}}48^{\mathrm{m}}15\fs2261$, 
$\delta_{\mathrm{c1}} = 17\degr46\arcmin26\farcs469$ and 
$\alpha_{\mathrm{c2}} = 12^{\mathrm{h}}48^{\mathrm{m}}15\fs2258$, 
$\delta_{\mathrm{c2}} = 17\degr46\arcmin26\farcs468$, where the subscripts correspond 
to the labels in Table \ref{tab:ep096abc_param} and Figure \ref{fig:ep096abc_img}. The 
position uncertainty is less than 1\,mas, even after taking into account the random 
errors in the position of the phase reference calibrator. The peak of the elongated 
component (c2) is at a projected distance of 4.3\,mas or 1.9\,pc ($5.8\times10^{18}$\,cm)
from the peak of component c1.

In the process of phase-referencing, solving for residual phase errors determined 
for the calibrator source at a slightly different sky position and at slightly 
different times than the target observations, some random and/or systematic phase 
errors may have been left uncalibrated. As a result, spurious, low-level structures 
may appear in the maps. ASASSN-14li was too weak to apply self-calibration in phase 
and account for such errors. This problem can also be alleviated with the use of an 
additional source brighter than the target and observed in the same way, as done by 
\citet{yang16} in their observations of {\it Swift}\,J$164449.3+573451$. If the same 
structure is observed toward both the target and the check source, then the common 
structure can be attributed to errors in the phase calibration. In this case, our 
observations did not include such source, and only future observations can verify 
the morphology of the target. However, we did thoroughly test the reliability of 
the structures we find. We experimented with different weighting regimes while 
imaging to test whether the observed structure could be an artifact. There are, 
for example, no correlations between the sidelobes and the emitting components. 
We made images both with {\sc aips} and the Caltech imaging program {\sc difmap} 
\citep{shepherd95}, and consistently find the same source structures and intensities 
in all our imaging trials in both packages.

\floattable
\begin{deluxetable}{cccccccc}
\tablecaption{Source parameters. \label{tab:ep096abc_param}}
\tablecolumns{8}
\tablewidth{0pt}
\tablehead{
\colhead{Segment} & \colhead{FWHM}   & \colhead{rms}   &   \colhead{$P_{\nu}$} & \colhead{$S_{\nu}$} & \colhead{$L_{\nu}$ }  &  \colhead{Deconvolved}  & \colhead{log$_{10}$\tb{}} \\
\colhead{} & \colhead{(mas$^2$, deg)}  & \colhead{(\mujyb{})} & \colhead{(\mjyb{})} & \colhead{(\mjy{})}  & \colhead{($10^{27}$\,\lunits{})} & \colhead{size (mas$^2$, deg)} & \colhead{(K)}
}
\startdata
A & $14.5\times7.3$, 83 &  34.4 & $1.12\pm0.07$ & $1.12\pm0.07$ & $10.94\pm0.64$ & $3.8\times \cdots$, 105.3 & $>6.67$ \\
B & $11.5\times6.2$, 78 &  14.5 & $0.71\pm0.05$ & $0.72\pm0.05$ & ~$7.07\pm0.51$  & $3.8\times1.8$, ~67.6      & $6.70\pm0.03$ \\
C & $11.5\times6.2$, 78 &  40.0 & $1.54\pm0.13$ & $1.62\pm0.13$ & $15.84\pm1.26$ & $4.1\times1.5$, 144.6     & $8.06\pm0.03$ \\
\hline
\vspace{-0.2cm}B\tablenotemark{c1} & ~                   &  ~   & ~$0.47\pm0.03$ & $0.63\pm0.06$ & $6.14\pm0.63$  & $1.5\times1.4$, 111.2 & $7.16\pm0.04$  \\ \vspace{-0.2cm}
~                                 & $2.7\times2.4$, 71  & 13.6 &  ~            & ~             &  ~             &  ~             &  ~ \\
B\tablenotemark{c2}                & ~                   &  ~   & ~$0.09\pm0.01$ & $0.10\pm0.02$ & $0.97\pm0.22$  & $3.4\times1.6$, ~89.1 & $5.95\pm0.10$  \\
\enddata
\tablecomments{~Col. 1 -- major and minor synthesized beam fitted FWHM of each map. Col. 2 -- rms
 value. Col. 3 -- peak intensity.  Col. 4 -- flux density.  Col. 5 -- luminosity at the central
 frequency.  Col. 6 --  deconvolved 
size.  Col. 7 -- brightness temperature.}
\tablenotetext{c1}{~~~Core-like component} 
\tablenotetext{c2}{~~~Elongated component}
\end{deluxetable}

\section{The nature of the radio emission} \label{sec:nat}

\citet{alexander16} reported Very Large Array (VLA) observations of ASASSN-14li at 
similar dates and frequencies to those we report here. The fluxes they find on 2015 
April 21 ($2.13\pm0.09$\,\mjy{} at 1.8\,GHz) and 2015 June 17 ($2.24\pm0.06$\,\mjy{} 
at 1.8\,GHz and $1.26\pm0.04$\,\mjy{} at 5.0\,GHz) are significantly higher than we 
found, although the implied spectral index of $\alpha\simeq-0.6$ is similar. Effectively, 
the flux density at 1.8\,GHz from the VLA observations can be considered as constant within 
the uncertainties. In the case of the EVN measurements we observe an increase of $\sim30$\% 
at a similar frequency, although the significance of this apparent increase is low ($\sim3.5$ 
times the standard error of the mean). For all observations, the flux density at 5.0\,GHz is 
always lower than the corresponding flux at 1.7\,GHz (or 1.8\,GHz) in the same epoch.

It is known that the flux density recovered in phase-referenced VLBI experiments 
decreases with increasing observed frequency and calibrator-to-target separation 
\citep[e.g.,][]{martividal10}. For the frequencies we observed here and given the 
small separation between ASASSN-14li and J1250$+$1621, the loss in flux density 
should be of only a few percent, and thus not significant to explain the drop in 
flux density in the EVN measurements with respect to the VLA ones. Instead, we assume 
that there must be a significant amount of diffuse, extended emission recovered by the 
VLA B/BnA configuration beam ($\sim4\farcs3$). This could also explain the slightly 
higher flux ($2.6\pm0.4$\,\mjy{}) we measure in the FWHM$=5\farcs4\times5\farcs4$ map 
from the Faint Images of the Radio Sky at Twenty-cm survey \citep[FIRST;][]{becker95}, 
and the apparently even higher flux density ($3.2\pm1.4$\,\mjy{}) we measure in the 
FWHM$=45\arcsec\times45\arcsec$ map of the NRAO VLA Sky Survey \citep[NVSS;][]{condon98}. 
The 1.4\,GHz observations reported by \citet{vanvelzen16} at a resolution of 
11\arcsec--13\arcsec, are consistent with the flux densities measured in both the NVSS 
and FIRST images within the uncertainties. Given the different beam sizes, there is no 
strong evidence for missing flux when comparing pre- and post-TDE measurements. The 
diffuse emission we are resolving out with the EVN at both 1.7 and 5.0\,GHz, is likely 
related with the pre-TDE steady source, a putative AGN, as proposed by \citet{alexander16}
and \citet{vanvelzen16}.

\citet{prieto16} present evidence for a pre-existing AGN in PGC\,043234 based on the 
emission line ratios of diffuse gas near the galaxy. As noted above, \citet{alexander16} 
argue that there must be a steady source within the radio emission measured with the VLA
whose flux density follows $S_{\nu} \approx S_0 (\nu/1.4 \mathrm{GHz})^{-1}$. Such a steep
power-law for the flux density implies large radiative losses and an old radiative age 
for the putative AGN \citep[older than a few times $10^7$\,yr, e.g.,][]{murgia11}. The 
optically-thin spectral behavior is consistent with the non-thermal ($\tb{}>10^6$\,K) 
core-like component observed with the EVN. The spectral index is probably less steep than 
$-1$ as we infer from our EVN observations, where the spectral index is dominated by the 
core-like component, given that the elongated, fainter one represents only $<14$\% of the 
total flux density. 

\citet{holoien16} obtained a pre-TDE {\it ROSAT} X-ray flux limit for the AGN in the host. 
The corresponding hard X-ray (2--10\,keV) luminosity is $<1.27\times10^{41}$\,erg\,s$^{-1}$,
obtained using the Mission Count Rate Simulator at 
\url{http://heasarc.gsfc.nasa.gov/cgi-bin/Tools/w3pimms/w3pimms.pl} with an input photon 
index of 2 fitted in the 0.08--2.9\,keV energy range \citep{holoien16}. Considering the 
2--10\,keV luminosity and the 5\,GHz luminosity of the steady component described by 
\citet{alexander16}, we obtain $\mathrm{log}_{10}R_{\mathrm{X}}>-3.7$ for the radio to 
hard X-ray luminosity ratio \cite[as described by][]{terashima03}. The AGN thus lies in 
the region dominated by low-luminosity AGNs and hard-state Seyferts \citep[in agreement 
with][]{prieto16}, and we infer that the pre-TDE AGN was not very active, likely due to 
starvation \citep{ho02}. 

The origin of the possibly elongated component is less clear. The projected distance 
between the peak positions of the components is 1.9\,pc. If the elongated component
is an outflow or a jet related to ASASSN-14li, and was ejected on 2014 August 11--25
\citep{alexander16}, then its apparent velocity is ($v_{\mathrm{app}}$) is $7.4c$--$7.8c$. 
Such superluminal motion disagrees with the interpretation of ASASSN-14li as a 
non-relativistic outflow/jet or unbound debris \citep{alexander16,vanvelzen16,krolik16}.
Following \citet{bottcher12}, we calculate a lower limit for the bulk Lorentz factor, 
$\Gamma$, of 7.5--7.8, and a maximum viewing angle of 14\fdeg7--15\fdeg4 at which 
$v_{\mathrm{app}}$ can be achieved. Given that $\Gamma \sim v_{\mathrm{app}}$, 
the maximum viewing angle should be close to the critical angle, which in this case
is within the lower-end of the values typically found for steep-spectrum radio quasars
and the upper end of the values for flat-spectrum radio quasars \citep{urry95,padovani07}. 
We note, however, that the typical brightness temperature for jet features observed with VLBI 
is usually a few times $10^{11}$\,K \citep{kellermann07}. Those authors also found that 
there are no low-luminosity sources (as we find here) with fast apparent velocities. This
makes the relativistic jet/outflow scenario less plausible.

If the elongated component is a non- or mildly-relativistic outflow or jet from a past AGN flare 
or a previous TDE, then it would have been ejected before 2009, for an apparent velocity of $<1c$. 
In particular, if it is the unbound debris ejected from a past TDE, theoretical models predict
that the typical velocity would be  $\sim0.03c$ \citep{chen16, guillochon16}. With such a velocity, 
the unbound debris stream could traverse typically a distance of $\sim10$\,pc before being completely 
stalled by the interstellar medium \citep[ISM;][]{guillochon16}. Therefore, the elongated component 
which we detected at a distance of about 2\,pc from the core can be interpreted as an unbound debris 
stream still evolving in the free-expansion phase. Given the $\sim0.03c$ theoretical velocity and 
the projected distance between components c1 and c2, we also infer that the past TDE happened about 
150 years ago and the TDE rate in this post-starburst galaxy is about $5\times10^{-3}$\,yr$^{-1}$. 
Such a high event rate is not unusual for systems containing binary SMBHs \citep{chen09,chen11} and 
has been inferred for TDEs hosted in post-starburst galaxies \citep{arcavi14,french16}. Given this 
rate, we calculate the luminosity at 5\,GHz using our model of shock heating and synchrotron cooling 
of unbound debris assuming an ISM similar to that in the Galactic Center, and we find 
$\nu L_\nu \sim10^{35}$--$10^{37}$\,erg\,s$^{-1}$ \citep[figure 9 in][]{guillochon16}. This luminosity 
is also consistent with what we derived for the elongated component in Table \ref{tab:ep096abc_param}.

The host of ASASSN-14li is a post-starburst galaxy and has a morphology suggestive of 
previous merger activity \citep{prieto16}, where the presence of a binary BH would be 
expected. Thus, if components c1 and c2 are truly separate components, an alternative 
scenario for their origin would be a binary BH system. For a primary BH mass within 
$10^6$--$10^7$\,\msun{}, the sphere of gravitational influence is on the order of 
$10^{18}$--$10^{19}$\,cm. This is comparable to the observed separation and is on these 
scales where the TDE rate enhancement is expected to be significant 
\citep[e.g.,][]{chen09,chen11,liu13}.

\section{Conclusions} \label{sec:sum}

Milliarcsec resolution observations of ASASSN-14li and its host have allowed us to 
resolve the radio emission into two components connected by a bridge of diffuse 
emission. Due to their morphology, one of the sources could represent a putative 
AGN, whilst the possibly elongated source could correspond to a jet/outflow knot 
related to ASASSN-14li. If the proper motion of the elongated component is sustained  
at a rate of 5.2--5.4\,mas\,yr$^{-1}$, our future EVN observations (project ER045) 
should easily detect these changes. This scenario will favor the interpretation of 
ASASSN-14li as the first superluminal TDE jet ever resolved, although we note that 
superluminous jets are much brighter than what we find here. If no proper motion is 
noticeable in the new observations, this would imply that the elongated component 
moves indeed at subluminal speeds, hence favoring its interpretation as a past AGN/TDE 
flare. However, observations with a better {\it uv}-sampling are needed in order to 
corroborate the morphology of the fainter source (elongated vs. compact). If it turns out
to be compact, then there would be a higher possibility that this system is a binary BH, 
the one with the smallest separation ever found.

\acknowledgments
The authors are grateful to Zsolt Paragi for very useful discussions and for providing 
comments on our manuscript. We also thank Ran Wang for useful discussions and the anonymous 
referee for his/her comments and positive feedback. We acknowledge support by the Ministry 
of Economy, Development, and Tourism's Millennium Science Initiative through grant IC120009, 
awarded to The Millennium Institute of Astrophysics, MAS, Chile (C.R.-C., J.L.P.), and from 
CONICYT through FONDECYT grants 3150238 (C.R.-C.) and 1151445 (J.L.P.). X.C. acknowledges 
support by the China-Conicyt Fellowship (CAS15002). S.D. is supported by ``the Strategic 
Priority Research Program --The Emergence of Cosmological Structures'' of the Chinese 
Academy of Sciences (Grant No. XDB09000000) and Project 11573003 supported by NSFC. C.S.K. 
and K.Z.S. are supported by NSF grants AST-1515876 and AST-1515927. T.W.-S.H. is supported 
by the DOE Computational Science Graduate Fellowship, grant number DE-FG02-97ER25308. The 
European VLBI Network is a joint facility of independent European, African, Asian, and North 
American radio astronomy institutes. Scientific results from data presented in this publication 
are derived from the EVN project code EP096. The research leading to these results has received 
funding from the European Commission Seventh Framework Programme (FP/2007-2013) under grant 
agreement No. 283393 (RadioNet3).

\vspace{5mm}

\software{{\sc aips}, {\sc difmap}, Python}


\begin{thebibliography}{}
\bibitem[Alexander et al.(2016)]{alexander16} Alexander, K.~D., Berger, E., Guillochon, J., Zauderer, B.~A., \& Williams, P.~K.~G.\ 2016, \apjl, 819, L25 
\bibitem[Arcavi et al.(2014)]{arcavi14} Arcavi, I., Gal-Yam, A., Sullivan, M., et al.\ 2014, \apj, 793, 38 
\bibitem[Becker et al.(1995)]{becker95} Becker, R.~H., White, R.~L., \& Helfand, D.~J.\ 1995, \apj, 450, 559 
\bibitem[Berger et al.(2012)]{berger12} Berger, E., Zauderer, A., Pooley, G.~G., et al.\ 2012, \apj, 748, 36 
\bibitem[Bloom et al.(2011)]{bloom11} Bloom, J.~S., Giannios, D., Metzger, B.~D., et al.\ 2011, Science, 333, 203 
\bibitem[Boettcher et al.(2012)]{bottcher12} Boettcher, M., Harris, D.~E., \& Krawczynski, H.\ 2012, Relativistic Jets from Active Galactic Nuclei, by M.~Boettcher, D.E.~Harris, ahd H.~Krawczynski, 425 pages.~ Berlin: Wiley, 2012,  
\bibitem[Bower et al.(2013)]{bower13} Bower, G.~C., Metzger, B.~D., Cenko, S.~B., Silverman, J.~M., \& Bloom, J.~S.\ 2013, \apj, 763, 84 
\bibitem[Brown et al.(2016)]{brown16} Brown, J.~S., W.-S Holoien, T., Auchettl, K., et al.\ 2016, arXiv:1609.04403 
\bibitem[Burrows et al.(2011)]{burrows11} Burrows, D.~N., Kennea, J.~A., Ghisellini, G., et al.\ 2011, \nat, 476, 421 
\bibitem[Cenko et al.(2016)]{cenko16} Cenko, S.~B., Cucchiara, A., Roth, N., et al.\ 2016, \apjl, 818, L32 
\bibitem[Chen et al.(2009)]{chen09} Chen, X., Madau, P., Sesana, A., \& Liu, F.~K.\ 2009, \apjl, 697, L149 
\bibitem[Chen et al.(2011)]{chen11} Chen, X., Sesana, A., Madau, P., \& Liu, F.~K.\ 2011, \apj, 729, 13 
\bibitem[Chen et al.(2016)]{chen16} Chen, X., G{\'o}mez-Vargas, G.~A., \& Guillochon, J.\ 2016, \mnras, 458, 3314 
\bibitem[Condon et al.(1998)]{condon98} Condon, J.~J., Cotton, W.~D., Greisen, E.~W., et al.\ 1998, \aj, 115, 1693 
\bibitem[Evans \& Kochanek(1989)]{evans89} Evans, C.~R., \& Kochanek, C.~S.\ 1989, \apjl, 346, L13 
\bibitem[Fomalont(2005)]{fomalont05} Fomalont, E.~B.\ 2005, Astrometry in the Age of the Next Generation of Large Telescopes, 338, 335 
\bibitem[French et al.(2016)]{french16} French, K.~D., Arcavi, I., \& Zabludoff, A.\ 2016, \apjl, 818, L21 
\bibitem[Generozov et al.(2017)]{generozov16} Generozov, A., Mimica, P., Metzger, B.~D., et al.\ 2017, \mnras, 464, 2481 
\bibitem[Guillochon et al.(2016)]{guillochon16} Guillochon, J., McCourt, M., Chen, X., Johnson, M.~D., \& Berger, E.\ 2016, \apj, 822, 48 
\bibitem[Hills(1975)]{hills75} Hills, J.~G.\ 1975, \nat, 254, 295 
\bibitem[Holoien et al.(2016)]{holoien16} Holoien, T.~W.-S., Kochanek, C.~S., Prieto, J.~L., et al.\ 2016, \mnras, 455, 2918
\bibitem[Ho(2002)]{ho02} Ho, L.~C.\ 2002, \apj, 564, 120 
\bibitem[Irwin et al.(2015)]{irwin15} Irwin, J.~A., Henriksen, R.~N., Krause, M., et al.\ 2015, \apj, 809, 172 
\bibitem[Jiang et al.(2016)]{jiang16} Jiang, N., Dou, L., Wang, T., et al.\ 2016, \apjl, 828, L14 
\bibitem[Kellermann et al.(2007)]{kellermann07} Kellermann, K.~I., Kovalev, Y.~Y., Lister, M.~L., et al.\ 2007, \apss, 311, 231 
\bibitem[Komossa(2015)]{komossa15} Komossa, S.\ 2015, Journal of High Energy Astrophysics, 7, 148 
\bibitem[Kormendy \& Ho(2013)]{kormendy13} Kormendy, J., \& Ho, L.~C.\ 2013, \araa, 51, 511 
\bibitem[Krolik et al.(2016)]{krolik16} Krolik, J., Piran, T., Svirski, G., \& Cheng, R.~M.\ 2016, \apj, 827, 127 
\bibitem[Levan et al.(2011)]{levan11} Levan, A.~J., Tanvir, N.~R., Cenko, S.~B., et al.\ 2011, Science, 333, 199 
\bibitem[Liu \& Chen(2013)]{liu13} Liu, F.~K., \& Chen, X.\ 2013, \apj, 767, 18 
\bibitem[Mart{\'{\i}}-Vidal et al.(2010)]{martividal10} Mart{\'{\i}}-Vidal, I., Ros, E., P{\'e}rez-Torres, M.~A., et al.\ 2010, \aap, 515, A53 
\bibitem[Miller et al.(2015)]{miller15} Miller, J.~M., Kaastra, J.~S., Miller, M.~C., et al.\ 2015, \nat, 526, 542 
\bibitem[Murgia et al.(2011)]{murgia11} Murgia, M., Parma, P., Mack, K.-H., et al.\ 2011, \aap, 526, A148 
\bibitem[Padovani(2007)]{padovani07} Padovani, P.\ 2007, The First GLAST Symposium, 921, 19 
\bibitem[Prieto et al.(2016)]{prieto16} Prieto, J.~L., Kr{\"u}hler, T., Anderson, J.~P., et al.\ 2016, \apjl, 830, L32 
\bibitem[Rees(1988)]{rees88} Rees, M.~J.\ 1988, \nat, 333, 523 
\bibitem[Shappee et al.(2014)]{shappee14} Shappee, B.~J., Prieto, J.~L., Grupe, D., et al.\ 2014, \apj, 788, 48 
\bibitem[Shepherd et al.(1995)]{shepherd95} Shepherd, M.~C., Pearson, T.~J., \& Taylor, G.~B.\ 1995, \baas, 27, 903 
\bibitem[Shields \& Wheeler(1978)]{shields78} Shields, G.~A., \& Wheeler, J.~C.\ 1978, \apj, 222, 667 
\bibitem[Shull(1983)]{shull83} Shull, J.~M.\ 1983, \apj, 264, 446 
\bibitem[Tremblay et al.(2016)]{tremblay16} Tremblay, G.~R., Oonk, J.~B.~R., Combes, F., et al.\ 2016, \nat, 534, 218 
\bibitem[van Velzen et al.(2013)]{vanvelzen13} van Velzen, S., Frail, D.~A., K{\"o}rding, E., \& Falcke, H.\ 2013, \aap, 552, A5 
\bibitem[van Velzen et al.(2016)]{vanvelzen16} van Velzen, S., Anderson, G.~E., Stone, N.~C., et al.\ 2016, Science, 351, 62 
\bibitem[Terashima \& Wilson(2003)]{terashima03} Terashima, Y., \& Wilson, A.~S.\ 2003, \apj, 583, 145 
\bibitem[Urry \& Padovani(1995)]{urry95} Urry, C.~M., \& Padovani, P.\ 1995, \pasp, 107, 803 
\bibitem[Yang et al.(2016)]{yang16} Yang, J., Paragi, Z., van der Horst, A.~J., et al.\ 2016, \mnras, 462, L66  
\bibitem[Zauderer et al.(2011)]{zauderer11} Zauderer, B.~A., Berger, E., Soderberg, A.~M., et al.\ 2011, \nat, 476, 425 
\end{thebibliography}
\end{document}